\documentclass[twocolumn]{aastex631}

\usepackage[hyphens]{url}
\usepackage{multirow}

\begin{document}

\title{Three Brown Dwarfs Masquerading as High-Redshift Galaxies in JWST Observations}

\correspondingauthor{Shu Wang}
\email{shuwang@nao.cas.cn}

\author[0009-0000-7976-7383]{Zhijun Tu}
\affiliation{CAS Key Laboratory of Optical Astronomy, National Astronomical Observatories, Chinese Academy of Sciences, Beijing 100101, People's Republic of China}
\affiliation{School of Astronomy and Space Sciences, University of Chinese Academy of Sciences, Beijing 100049, People's Republic of China}

\author[0000-0003-4489-9794]{Shu Wang}
\affiliation{CAS Key Laboratory of Optical Astronomy, National Astronomical Observatories, Chinese Academy of Sciences, Beijing 100101, People's Republic of China}

\author[0000-0001-7084-0484]{Xiaodian Chen}
\affiliation{CAS Key Laboratory of Optical Astronomy, National Astronomical Observatories, Chinese Academy of Sciences, Beijing 100101, People's Republic of China}
\affiliation{School of Astronomy and Space Sciences, University of Chinese Academy of Sciences, Beijing 100049, People's Republic of China}
\affiliation{Institute for Frontiers in Astronomy and Astrophysics, Beijing Normal University, Beijing 102206, People's Republic of China}

\author{Jifeng Liu}
\affiliation{CAS Key Laboratory of Optical Astronomy, National Astronomical Observatories, Chinese Academy of Sciences, Beijing 100101, People's Republic of China}
\affiliation{School of Astronomy and Space Sciences, University of Chinese Academy of Sciences, Beijing 100049, People's Republic of China}
\affiliation{Institute for Frontiers in Astronomy and Astrophysics, Beijing Normal University, Beijing 102206, People's Republic of China}
\affiliation{New Cornerstone Science Laboratory, National Astronomical Observatories, Chinese Academy of Sciences, Beijing 100012, People's Republic of China}

\begin{abstract}

We report the spectroscopic identification of three brown dwarf candidates—--o005\_s41280, o006\_s00089, and o006\_s35616—--discovered in the RUBIES using James Webb Space Telescope (JWST) Near-Infrared Spectrograph (NIRSpec) PRISM/CLEAR spectroscopy. We fit these sources with multiple substellar atmosphere models and present the atmospheric parameters, including effective temperature ($T_\mathrm{eff}$), surface gravity, and other derived properties. The results suggest that o005\_s41280 and o006\_s35616, with $T_\mathrm{eff}$ in the ranges of 2100–-2300 K and 1800--2000 K, are likely L dwarfs, while o006\_s00089, with $T_\mathrm{eff} < 1000$ K, is consistent with a late T dwarf classification. The best-fit model spectra provide a reasonable match to the observed spectra. However, distinct residuals exist in the $Y$, $J$, and $H$ bands for the two L dwarf candidates, particularly for o006\_s35616. Incorporating the extinction parameter into the fitting process can significantly reduce these residuals. The distance estimates indicate that these candidates are about 2 kpc away. The analysis of the color-color diagram using multiple JWST NIRcam photometry suggests that cooler T dwarfs, such as o006\_s00089, overlap with little red dots (LRDs), while hotter L dwarfs, like o005\_s41280 and o006\_s35616, tend to contaminate the high-redshift galaxy cluster. These findings suggest a brown dwarf contamination rate of approximately 0.1\% in extragalactic deep field surveys, with L dwarfs being more frequently detected than cooler T and Y dwarfs.
\end{abstract}


\section{Introduction} \label{sec:intro}

The advent of the James Webb Space Telescope (JWST) has revolutionized our exploration of the distant universe, unveiling numerous red, compact sources that challenge our understanding of galaxy formation and evolution \citep[e.g.,][]{2024NatAs...8..126B,2024ApJ...963..128B}. Among these intriguing discoveries is a population of little red dots (LRDs), compact sources with extremely red colors initially identified in deep-field imaging surveys. These objects are hypothesized to represent faint and/or highly reddened active galactic nuclei (AGNs) at high redshifts (e.g., $z > 3$), with some hosting accreting supermassive black holes enshrouded by significant dust \citep{2024ApJ...963..129M}. Studies of LRDs not only provide insights into black hole formation and growth but also help constrain the role of AGN feedback in regulating star formation within their host galaxies.

To further explore this population, the JWST General Observer (GO) program 4233 (PI: A. de Graaff), named Red Unknowns: Bright Infrared Extragalactic Survey (RUBIES), targets the brightest F444W-selected sources from two key deep fields, CEERS \citep{2023ApJ...946L..13F} and PRIMER \citep{2021jwst.prop.1837D}, using JWST Near-Infrared Spectrograph (NIRSpec) multi-object spectroscopy \citep{2024arXiv240905948D}. This program spans 18 pointings spread across two legacy extragalactic deep fields ($\sim150\,\mathrm{arcmin^2}$), with the Galactic coordinates of CEERS field centered at \(l=96.38^\circ\), \(b=+59.42^\circ\), and the PRIMER field at \(l=169.83^\circ\), \(b=-60.00^\circ\) (see \citealt[][Figure 1]{2024arXiv240905948D}). It observes $\sim$3000 targets across $1 < z_\mathrm{phot} < 10$ with both the NIRSpec PRISM and G395M dispersers, and $\sim$1500 targets at $z_\mathrm{phot} > 3$ using only the G395M disperser. The primary goals are to refine redshift and mass estimates for bright $z > 7$ galaxies, reveal the nature of extremely red sources, and trace the star formation histories of galaxies at intermediate redshifts. With moderate exposure times and a combination of prism and grating modes, this survey is designed to provide a legacy dataset of a combination of low- and medium-resolution spectra, offering transformative insights into the nature of LRDs, high-redshift galaxy populations, interstellar medium properties, dust, and active black holes \citep{2024arXiv240302304W,2024ApJ...969L..13W,2024arXiv240903829W}.

However, contaminants objects, such as brown dwarfs, are possibly present within the RUBIES due to their spectral and photometric similarity to LRDs or high-redshift galaxies in the near-infrared bands \citep[e.g.,][]{2023ApJ...957L..27L,2024ApJ...962..177B}. Brown dwarfs, particularly those of late T and Y spectral types, exhibit V-shaped spectral energy distributions (SEDs) with pronounced features near 1 and 4 $\mu$m, driven by molecular absorption bands of water and methane \citep[e.g.,][]{2005ARA&A..43..195K,2014ASInC..11....7B}. These features can closely resemble the rest-frame optical and ultraviolet breaks observed in LRDs or high-redshift galaxies, especially those heavily obscured by dust. Furthermore, brown dwarfs share photometric properties with the point-like sources selected in extragalactic deep field surveys, including similar colors (e.g., F115W$-$F277W and F277W$-$F444W) and bright magnitude in F444W \citep[][]{2024ApJ...964...66H}. This overlap in SED characteristics and photometric profiles makes them natural contaminants in color-selected samples of LRDs and high-redshift galaxies.

In the specific case of RUBIES, the focus on bright F444W-selected sources introduces brown dwarf contamination. The wide-area coverage of the fields---CEERS and PRIMER, further enhances the chance of encountering brown dwarfs \citep[e.g.,][]{2023MNRAS.523.4534W,2024ApJ...964...66H}. Thus, while the proposal primarily aims to uncover distant, high-redshift galaxies and LRDs, the inclusion of brown dwarfs highlights the importance of spectroscopic follow-up for robust source classification.

In this study, we present the spectroscopic identification of three brown dwarf candidates within the RUBIES. We describe the data we used and preliminary sample screening in Section~\ref{sec:sample}. In Section~\ref{sec:methods}, we introduce the atmospheric models we used, and the Bayesian framework with nested sampling employed to analyze their physical parameters and properties. Subsequently, the fitting results and their analysis are provided in Section~\ref{sec:results}. In Section~\ref{sec:discussion}, we examine the observed colors of the three brown dwarf candidates, highlighting their distributions relative to LRDs and high-redshift galaxies. Finally, we summarize our findings and conclusions in Section~\ref{sec:conclusion}.

\section{Data and Sample} \label{sec:sample}

We retrieved all publicly available stage 3 NIRSpec PRISM/CLEAR spectra of the RUBIES (GO program 4233) from the Mikulski Archive for Space Telescopes (MAST)\footnote{\url{https://mast.stsci.edu/search/ui/\#/jwst}}. The stage 3 spectra are combined, calibrated science products. The NIRSpec PRISM/CLEAR provides low-resolution spectroscopy ($R\sim100$) across the whole near-infrared wavelength range (0.6--5.3 $\mu$m), which is sufficient to characterize the features of brown dwarfs in $\sim1$ and $\sim4\,\mu$m. 

In total, we derived 3194 spectra of sources. These stage 3 spectra were processed routinely by MAST using the official JWST Science Calibration Pipeline (Version 1.14.0), along with the Calibration Reference Data System (CRDS) version 11.17.19 under the operational context file 1256.pmap. This pipeline applies standard calibrations, including background subtraction, flat-fielding, flux calibration, and spectral extraction, resulting in fully reduced 1D spectra suitable for scientific analysis.
 
To identify brown dwarf candidates, we visually inspected all the 3194 NIRSpec PRISM/CLEAR spectra obtained from RUBIES for the characteristic peak features in the 1--2.4 $\mu$m, and compared the near-infrared spectra with standard spectra for brown dwarfs \citep{2017AJ....153...46L,2024ApJ...976...82T}. This distinct pattern observed in brown dwarfs, includes pronounced absorption features caused by the molecules of methane and water.

Sources exhibiting this peak signature were selected for further analysis, as this feature strongly suggests the presence of substellar objects with cool atmospheres. After a thorough manual inspection, we found three sources that the spectra exhibited features characteristic of brown dwarfs.

The essential information and photometry for the three sources are provided in Table~\ref{tab:BD_info}. Their object IDs are selected from their observation numbers and source numbers in RUBIES (e.g., o006\_s00089 from observation number 6 and source number 00089). The photometric values of the JWST NIRCam filters were obtained by cross-matching our sources with the photometry catalog of CEERS from \citet{2024AA...691A.240M}, as all three sources belong to the CEERS field.

\section{Characterizing the Brown Dwarf Candidates} \label{sec:methods}

\begin{deluxetable*}{cccccccccc}
\tablecaption{Fundamental Information and Photometry of Brown Dwarf Candidates\label{tab:BD_info}}
\setlength{\tabcolsep}{0.35cm}
\tablehead{\colhead{Object ID} & \colhead{R.A.} & \colhead{Decl.} & \colhead{F115W} & \colhead{F150W} & \colhead{F200W} & \colhead{F277W} & \colhead{F356W} & \colhead{F410M} & \colhead{F444W} \\ 
\colhead{} & \colhead{(J2000)} & \colhead{(J2000)} & \colhead{($\mu$Jy)} & \colhead{($\mu$Jy)} & \colhead{($\mu$Jy)} & \colhead{($\mu$Jy)} & \colhead{($\mu$Jy)} & \colhead{($\mu$Jy)} & \colhead{($\mu$Jy)} } 
\startdata
o005\_s41280 & 214.828485 & 52.810830 & 0.22364 & 0.22568 & 0.23319 & 0.15441 & 0.20739 & 0.21956 & 0.1821 \\
o006\_s00089 & 214.910293 & 52.860062 & 0.07109 & 0.03789 & 0.02717 & 0.01328 & 0.04004 & 0.14769 & 0.11403 \\
o006\_s35616 & 214.938850 & 52.873855 & 0.93326 & 1.30806 & 1.51984 & 1.07989 & 1.38642 & 1.34722 & 1.13382 \\
\enddata
\tablecomments{All the photometric values of the JWST NIRcam filters are taken from \citet{2024AA...691A.240M}. These values are in the units of $\mu$Jy with zero-point $Z=23.9$.}
\end{deluxetable*}

In this section, we will introduce the substellar atmospheric models utilized in this study, the methodology used to analyze the spectra of the three brown dwarf candidates, and describe the processing applied to these observed spectra.

\subsection{Atmospheric Models}
For all brown dwarf candidates, we first utilized the Sonora Elf Owl atmospheric model \citep{Mukherjee2024ApJ} to fit with observed spectra and derive key atmospheric parameters. The Sonora Elf Owl is a radiative-convective equilibrium (RCE) cloudless atmospheric model that incorporate disequilibrium chemistry. Disequilibrium chemistry, driven by vertical mixing, is a well-documented feature in brown dwarf atmospheres \citep{Leggett2015ApJ, Leggett2017ApJ}. Moreover, cloudless models have been shown to better match the atmospheres of T and Y dwarfs \citep[e.g.,][]{Tremblin2015ApJ}. The Sonora Elf Owl model is the latest generation of the Sonora model series, featuring a comprehensive parameter space that includes:
\begin{itemize}
    \item Effective temperature: $T_\mathrm{eff} = 275 - 2400\,\mathrm{K}$;
    \item Surface gravity: $\log g = 3.25 - 5.5\,[\mathrm{cm\,s^{-2}}]$;
    \item Metallicity: $[\mathrm{M/H}] = -1, -0.5, 0, 0.5, 0.7, 1\,\mathrm{dex}$;
    \item Vertical eddy diffusion coefficient: $\log K_\mathrm{zz} = 2, 4, 7, 8, 9\,[\mathrm{cm^2\,s^{-1}}]$;
    \item Carbon-to-oxygen ratio: $\mathrm{C/O} = 0.22, 0.458, 0.687, 1.12$ ($\mathrm{C/O_\odot = 0.458}$, \citealt{Lodders2009LanB}).
\end{itemize}
With a total of 43200 unique model grids, the Sonora Elf Owl offers extensive parameter coverage, making it appropriate for detailed studies of specific cases. 

In addition to the Sonora Elf Owl atmospheric model, other models were also utilized to cross-validate and enhance the consistency of the derived physical parameters. Once the initial atmospheric parameters were derived using the Sonora Elf Owl model, further spectral fitting was conducted using either the ATMO2020++ model ($250 - 1200\,\mathrm{K}$) or BT-settl CIFIST model ($1200 - 4000\,\mathrm{K}$). The choice of model for additional spectral fitting of each source was determined by the derived $T_\mathrm{eff}$ using the Sonora Elf Owl model. 

The ATMO2020++ model was applied to objects with $T_\mathrm{eff} \leq 1200\,\mathrm{K}$, as it is also an RCE cloudless atmospheric model with disequilibrium chemistry and optimized for the cooler atmospheres of late T and Y dwarfs \citep[e.g.,][]{2023ApJ...959...86L,2024ApJ...976...82T,2024arXiv241010939Z}. The ATMO2020++ model, a modified version of the ATMO2020 models \citep{Phillips2020A&A}, is designed specifically for late T and Y dwarfs, with an adjusted adiabatic index ($\gamma$) to better replicate the observed spectra of cooler brown dwarfs \citep{Meisner2023AJ}. The ATMO2020++ model includes:
\begin{itemize}
    \item Effective temperature: $T_\mathrm{eff} = 250 - 1200\,\mathrm{K}$,
    \item Surface gravity: $\log g = 2.5 - 5.5\,[\mathrm{cm\,s^{-2}}]$,
    \item Metallicity: $[\mathrm{M/H}] = -1, -0.5, 0, 0.3\,\mathrm{dex}$.
\end{itemize}

For sources with $T_\mathrm{eff} > 1200\,\mathrm{K}$, we utilized the BT-settl CIFIST model. This model accounts for clouds and dust in the atmospheres of warmer substellar objects and incorporates disequilibrium chemistry effects \citep{2012RSPTA.370.2765A,2015A&A...577A..42B}, as the spectra of hotter substellar objects may still be affected by the opacity of these clouds \citep{2012ApJ...756..172M}. The cloud substellar atmospheric models could match those of observed L dwarfs well than the cloudless atmospheric models \citep[e.g.,][]{2008ApJ...678.1372C,2024AJ....167..168M}. The parameter space of the BT-settl CIFIST model that we use:
\begin{itemize}
    \item Effective temperature: $T_\mathrm{eff} = 1200 - 4000\,\mathrm{K}$,
    \item Surface gravity: $\log g = 2.5 - 5.5\,[\mathrm{cm\,s^{-2}}]$,
\end{itemize}
with solar metallicity defined by \citet{2011SoPh..268..255C}. Compared to other versions of the BT-Settl models, the BT-settl CIFIST model leverages updated molecular line lists and improved convective mixing lengths based on multidimensional radiative hydrodynamic simulations \citep{2015A&A...577A..42B}.

By integrating these additional models into the analysis, we were able to validate the atmospheric parameters derived from these models and improve the accuracy of our physical interpretations, particularly for cases where the fitted parameter results in one of the models converge to the boundaries of that model.

\subsection{Fitting with Atmospheric Models} \label{fit method}
Forward-modeling analyses commonly involve comparing an observed spectrum with synthetic spectra generated from atmospheric models, with the best-fit parameters determined through least-squares minimization or Bayesian framework. Due to the typically coarse spacing of model grids, linear interpolation between synthetic spectra is often used to estimate parameters for values lying between grid points.

We began our analysis by linearly interpolating the three atmospheric models across their available parameter grids. The resulting interpolated model spectra were then convolved with a Gaussian profile based on the wavelength-resolution relation for the NIRSpec PRISM/CLEAR spectra, as described by \citet{2022A&A...661A..80J}, ensuring a consistent spectral resolution for fitting.

Given the extremely faint flux and the absence of prominent features in the 0.6–0.95 $\mu$m range of brown dwarf spectra, the spectral data in this wavelength region was excluded for all observed spectra. Additionally, any flux points with anomalously high or low values were removed to ensure the reliability of the spectra.

As the actual distances of the three sources are unknown, we utilized the 3D extinction map from \citet{2019ApJ...887...93G} to evaluate the strength of foreground interstellar extinction. The results indicate that foreground interstellar extinctions of the three sources are negligible across a distance range of 0 to 60 kpc. This is consistent with the fact that these sources are located in the extragalactic deep field of CEERS (\(l=96.38^\circ\), \(b=+59.42^\circ\)), which are far outside the Galactic Plane. Their positions minimize the impact of interstellar extinction, unlike the common observations of brown dwarfs conducted within the Galactic Plane where significant extinction is often a concern.

To derive the posterior probability distributions of the physical parameters, we utilized the nested sampling Monte Carlo algorithm "MLFriends" \citep{untranest2016, ultranest2019}, implemented through the open-source code \texttt{UltraNest} \citep{ultranest}. The likelihood function for our model is given by

\begin{equation}
\ln L = -0.5 \sum_i \frac{(F_i - \frac{R^2}{D^2} M_i)^2}{\sigma_i^2} - \ln\sqrt{2\pi \sigma_i^2},
\end{equation}
where the $F_i$, $M_i$, and $\sigma_i$ are the observed flux, model flux, and observed flux error, respectively. The scaling factor $R^2/D^2$, corresponding to the square of the ratio of radius to distance, is treated as a free parameter in the fitting process. All prior distributions were assumed to be uniform. The logarithmic scaling factor $\log (R^2/D^2)$ was included with a prior range of $-10$ to $-30$.

It should be noted that our analysis only accounts for uncertainties in the observed flux, without considering other potential sources of error, such as those introduced by linear interpolation. Linear interpolation can contribute non-negligible uncertainties to the fitting results, but it typically has a minimal effect on the central values of the derived parameters \citep{2021ApJ...921...95Z,2024ApJ...976...82T}. Consequently, the uncertainties derived using this method could be underestimated. However, this does not affect the overall reliability of the parameter trends, comparisons across models, or the qualitative conclusions drawn from the analysis.

\section{Results} \label{sec:results}
\begin{figure*}[t]
\centering 
\includegraphics[width=1\textwidth,height=0.3\textheight]{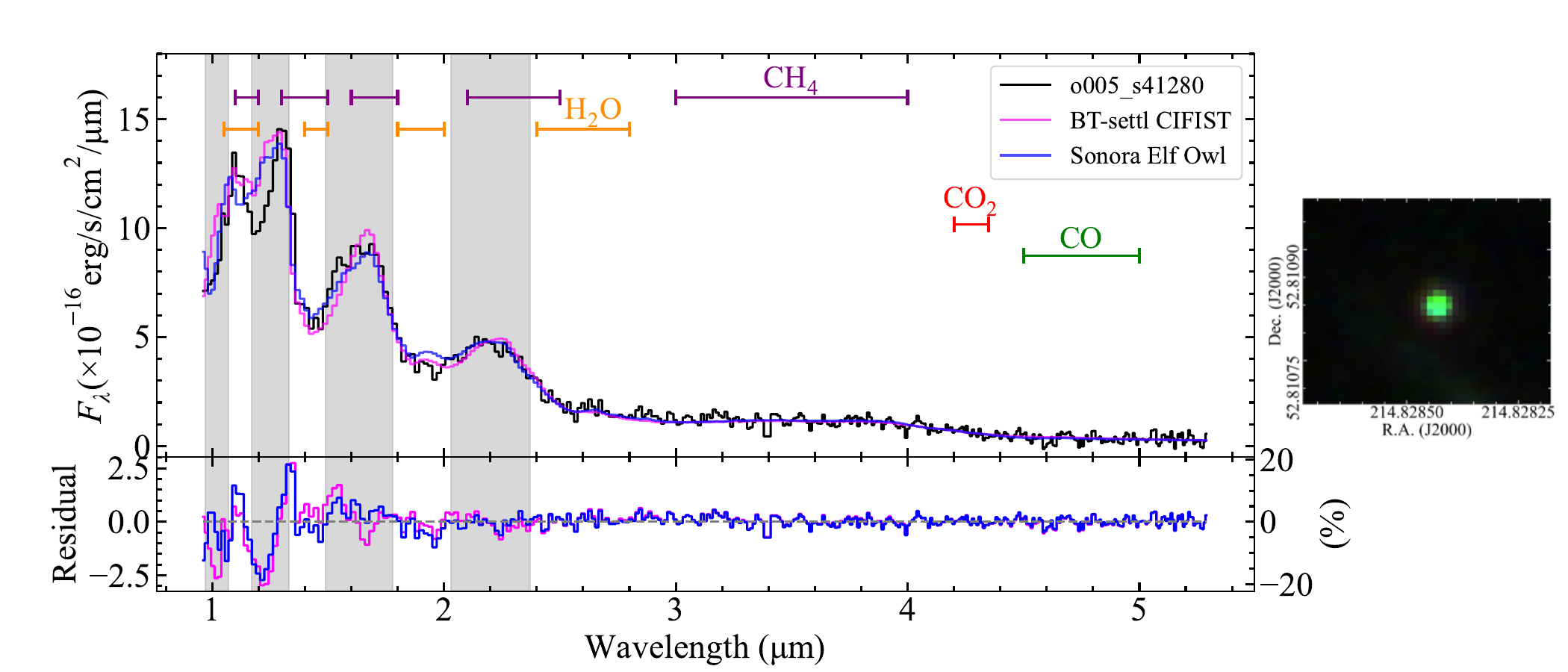}
\includegraphics[width=1\textwidth,height=0.3\textheight]{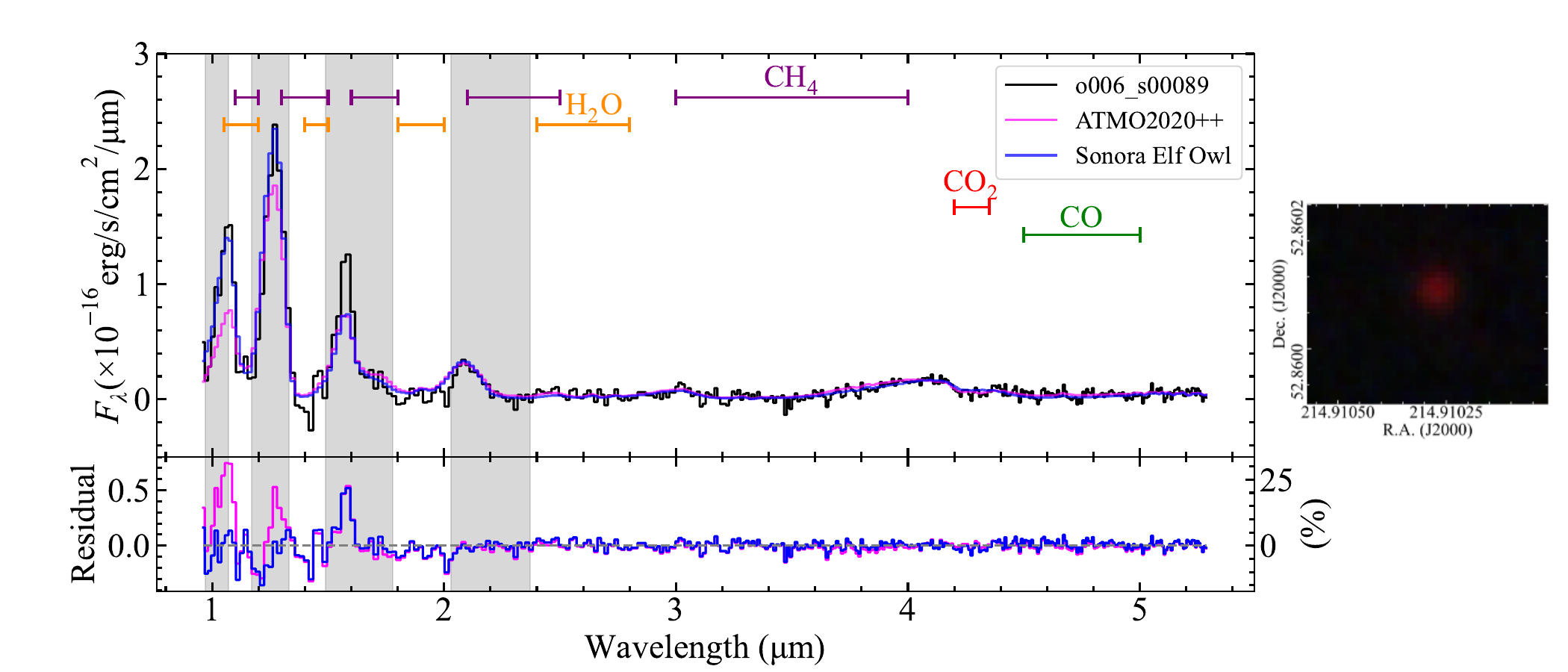}
\includegraphics[width=1\textwidth,height=0.3\textheight]{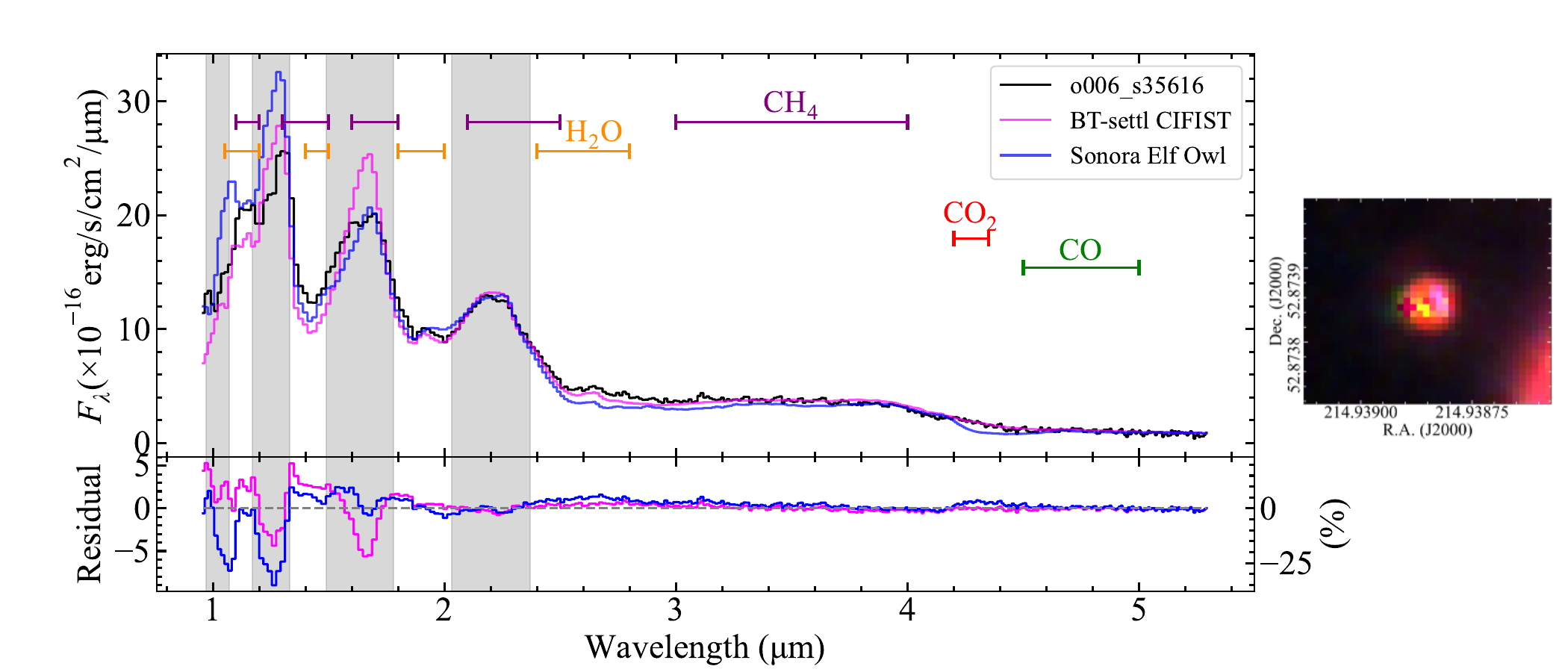}
\caption{The left panels compare the observed spectra of the three brown dwarf candidates with their corresponding best-fit model spectra, derived using the nested sampling Monte Carlo algorithm. The percentage residuals on the right axis are calculated by dividing the residuals by the corresponding maximum observed flux. The right panels display the RGB images of each source within a $2 \, \mathrm{arcsec}^2$ field of view, with red, green, and blue channels corresponding to F444W, F200W, and F150W photometry, respectively. The gray-shaded regions indicate the wavelength intervals corresponding to the $Y$, $J$, $H$, and $K$ photometric bands.\\\label{fig:MCMCfitting}} 
\end{figure*}

\begin{deluxetable*}{lcccccc}
\tablecaption{The Best-fit Parameters and Properties of the Three Brown Dwarf Candidates\label{tab:parameter}}
\setlength{\tabcolsep}{0.25cm}
\tablehead{\colhead{} & \multicolumn{2}{c}{o005\_s41280}  & \multicolumn{2}{c}{o006\_s00089} & \multicolumn{2}{c}{o006\_s35616}\\
\colhead{} & \colhead{Sonora Elf Owl} & \colhead{BT-settl CIFIST} & \colhead{Sonora Elf Owl} & \colhead{ATMO2020++} & \colhead{Sonora Elf Owl} & \colhead{BT-settl CIFIST}
} 
\startdata
$\log K_\mathrm{zz}$ [$\mathrm{cm^2\ s^{-1}}$] & $3.11^{+1.04}_{-0.77}$ & * & $3.53^{+0.56}_{-0.47}$ & * & $2.08^{+0.13}_{-0.06}$ & * \\
$T_\mathrm{eff}$ (K) & $2162.19^{+24.78}_{-21.52}$ & $2256.16^{+12.60}_{-12.19}$ & $955.41^{+15.63}_{-16.57}$ & $998.53^{+5.68}_{-7.05}$ & $1810.27^{+3.63}_{-3.54}$ & $1914.81^{+0.57}_{-0.59}$ \\
$\log g$ [$\mathrm{cm\ s^{-2}}$] & $3.29^{+0.06}_{-0.03}$ & $3.50^{+0.02}_{-0.03}$ & $4.34^{+0.18}_{-0.20}$ & $5.49\pm 0.01$ & $3.25\pm 0.01$ & $4.30\pm 0.01$ \\
$\mathrm{[M/H]}$ & $-0.18^{+0.07}_{-0.08}$ & * & $0.41\pm 0.05$ & $0.29\pm 0.01$ & $1.00\pm 0.01$ & * \\
C/O & $0.73\pm 0.01$ & * & $0.59\pm 0.03$ & * & $0.75\pm 0.01$ & * \\
$\log(R^2/D^2)$ & $-23.93\pm 0.02$ & $-24.01\pm 0.01$ & $-23.74\pm 0.03$ & $-23.84\pm 0.01$ & $-23.25\pm 0.01$ & $-23.37\pm 0.01$ \\
$M/M_\mathrm{Jup}$ & $<8.4$ & $<8.4$ & $11.5^{+4.0}_{-3.4}$ & $>66.7$ & $<8.4$ & $15.3^{+0.2}_{-0.3}$ \\
Age (Myr) &$<1$ &$<1$ & $104.8^{+119.9}_{-62.0}$ & $>10000$ & $<1$ & $44.9^{+0.4}_{-0.7}$ \\
$R/R_\mathrm{Jup}$ & $>2.3$ & $>2.3$ & $1.2\pm 0.1$ & $<0.8$ & $>2.0$ & $1.4\pm 0.1$ \\
Distance (kpc) & $>4.8$ & $>5.3$ & $2.0\pm 0.2$ & $<1.5$ & $>2.0$ & $1.6\pm 0.2$ \\
\enddata

\tablecomments{The fitting parameters and their uncertainties are derived from the marginalized posterior distributions, represented by the 16th, 50th, and 84th percentiles, corresponding to the median and the 68\% credible interval. The uncertainties in mass, age, and radius are derived from the values and associated uncertainties of the fitted parameters $T_\mathrm{eff}$ and $\log g$. Similarly, the uncertainties in distance are calculated using the values and uncertainties of the radius and the scaling factor $\log(R^2/D^2)$. In this table, uncertainties smaller than 0.01 have been rounded up to 0.01 for clarity. An asterisk (*) in the table indicates that this parameter does not exist in this model or that it is not a free parameter.}
\end{deluxetable*}

\subsection{Parameters and Physical Properties}

The fitting results for the three brown dwarf candidates are summarized in Table~\ref{tab:parameter}. Additionally, Figure~\ref{fig:MCMCfitting} presents comparisons between the observed spectra and the corresponding best-fit model spectra for each source. The figure also includes RGB images of each source within a $2 \, \mathrm{arcsec}^2$ field of view, where the red channel corresponds to F444W photometry, the green channel to F200W photometry, and the blue channel to F150W photometry. All photometric data were obtained from the JWST ERS program: 1345 (PI: S. Finkelstein).

We utilized the linearly interpolated evolutionary model of \citet{Chabrier2023AA} to estimate its mass, radius, and age based on the derived $T_\mathrm{eff}$ and $\log g$, which are also presented in Table~\ref{tab:parameter}. Once the radius $R$ was determined, the distance $D$ to each source was computed using the fitted scaling factor parameter, $\log (R^2/D^2)$, and the estimated radius. We note that, benefiting from the smaller flux uncertainties of o006\_s35616, the uncertainties in the most fitting parameters of o006\_s35616 are smaller than those of the other two sources.

The $T_\mathrm{eff}$ derived from the Sonora Elf Owl model for o006\_s00089 is $955.41^{+15.63}_{-16.57}\,\mathrm{K}$, which is below 1200 K. Consequently, this source was further analyzed using the ATMO2020++ model. For the other two sources, o005\_s41280 and o006\_s35616, $T_\mathrm{eff}=2162.19^{+24.78}_{-21.52}\,\mathrm{K}$ and $T_\mathrm{eff}=1810.27^{+3.63}_{-3.54}\,\mathrm{K}$, respectively, and thus we further used the BT-settl CIFIST model. The fitting results show that the $T_\mathrm{eff}$ obtained from the two models are generally consistent for each source, with differences typically within 100 K. Using the empirical relationship between $T_\mathrm{eff}$ and spectral type from \citet{2002ApJ...564..421B}, we estimate the spectral type ranges for each source. Specifically, o005\_s41280 is classified as approximately M9--L1 type, o006\_s00089 as T5--T7 type, and o006\_s35616 as L3--L4 type.

The $\log g$ values derived for the three sources vary between models, with the Sonora Elf Owl model consistently yielding lower $\log g$ values compared to other models. As shown in Figure~\ref{fig:MCMCfitting}, the best-fit model spectra from the two models for each source are quite similar and effectively match the observed spectra. This suggests that the differences in $\log g$ primarily arise from different model assumptions \citep{2024ApJ...976...82T}.

As the mass, age, and radius are derived from the fitted $T_\mathrm{eff}$ and $\log g$ values, discrepancies in $\log g$ lead to variations in the physical properties obtained from different models. Besides, it is important to note the limitations of the evolutionary model. For o005\_s41280, the points ($T_\mathrm{eff}, \log g$) derived from both models are outside the boundary of the evolutionary model grids due to extremely low $\log g$. Similarly, for o006\_s35616, the $\log g$ value obtained using the Sonora Elf Owl model falls beyond the grid boundaries of the evolutionary model at the corresponding $T_\mathrm{eff}$. A comparable issue arises for o006\_s00089 when analyzed with the ATMO2020++ model. As a result, the masses, ages, and radii derived under these scenarios are presented in Table~\ref{tab:parameter} as reference values only, acknowledging the limitations imposed by the model grid boundaries.

Alternatively, we estimated the ages of the three sources by calculating their vertical heights $Z$ above the Galactic plane and utilizing the relationship between $Z$ and stellar age provided by \citet{2013A&A...560A.109H}. All three sources have Galactic latitudes $b$ of approximately 60 degrees. This means that even at a distance of 1 kpc, their vertical heights above the Galactic plane would reach approximately $Z \approx 0.86 \, \mathrm{kpc}$. According to \citet{2013A&A...560A.109H}, stars younger than 1 Gyr are confined to heights below $Z < 0.5 \, \mathrm{kpc}$. Thus, the ages derived for the two L-type brown dwarf candidates, o005\_s41280 and o006\_s35616, based on their $T_\mathrm{eff}$, $\log g$, and evolutionary models, are likely significantly underestimated. This underestimation results from the extremely low value in $\log g$ derived by the model, which leads to an overestimation of radius and, consequently, inflated distance estimates. 

To obtain more reasonable results, we adopted an age range of 1--5 Gyr for each source. Using the fitted $T_\mathrm{eff}$ and this age estimate, we constrained the radii of o005\_s41280, o006\_s00089, and o006\_s35616 to the ranges $0.9-2.0 \, R_\mathrm{Jup}$, $0.8-0.9 \, R_\mathrm{Jup}$, and $0.8-1.1 \, R_\mathrm{Jup}$, respectively. With these radius estimates and the fitted values of $\log(R^2/D^2)$, we derived the distances for each source to be $1.9-4.7 \, \mathrm{kpc}$, $1.4-1.7 \, \mathrm{kpc}$, and $0.9-1.4 \, \mathrm{kpc}$, respectively. It is worth noting that these results show slight discrepancies compared to those presented in Table~\ref{tab:parameter}, particularly for o005\_s41280, which is influenced by its extremely low fitted $\log g$. Achieving more accurate distance constraints for these sources will require improved $\log g$ estimations and refined atmospheric and evolutionary models. Additionally, identifying a larger sample of brown dwarf candidates, particularly those at greater distances, could provide valuable data to further optimize and calibrate these models.

\subsection{Comparison of Observed and Model Spectra}

In this section, we compare the observed spectra of the three brown dwarf candidates with their corresponding best-fit model spectra for four photometric bands, as well as the overall match, highlighting the differences between the observed data and the models derived using the BT-settl CIFIST, ATMO2020++, and Sonora Elf Owl models.

\begin{itemize}
    \item $Y$ band (0.97--1.07 $\mu$m): Significant discrepancies are observed in the $Y$ band. For o006\_s00089, the flux derived using the ATMO2020++ model underestimates the observed flux, whereas the Sonora Elf Owl model shows better agreement with the observations. For o006\_s35616, the BT-settl CIFIST model predicts weaker flux than observed, while the Sonora Elf Owl model overestimates the flux. For o005\_s41280, both models overestimate the observed flux in this band.
    \item $J$ band (1.17--1.33 $\mu$m): For the cooler source o006\_s00089, the differences between the model and observed spectra in the $J$ band are relatively minor. In contrast, the $J$ band shows the most pronounced overestimation for the two warmer L-type brown dwarf candidates, o005\_s41280 and o006\_s35616. These discrepancies are likely due to limitations in the models' treatment of molecular opacities and cloud effects.
    \item $H$ band (1.49--1.78 $\mu$m): In this band, the models generally align well with the observed spectra for o005\_s41280 and o006\_s35616, except that the BT-settl CIFIST model overestimates the flux for o006\_s35616. For o006\_s00089, however, both models underestimate the observed flux.
    \item $K$ band (2.03--2.37 $\mu$m): The model spectra capture the overall observed flux trends in the $K$ band reasonably well across all three sources. While minor differences persist, they are less significant compared to those in the $Y$ and $J$ bands.
\end{itemize}

Beyond the four photometric bands discussed above, the observed spectra of all three sources exhibit generally good agreement with the model spectra in the redder wavelength range of 2.4–5.3 $\mu$m. However, it is worth noting that for o006\_s35616, the Sonora Elf Owl model underestimates the observed flux significantly in the 2.6–3 $\mu$m and 4.2–4.4 $\mu$m regions.

\subsection{Spectral Fitting with Extinction Parameter} \label{subsec:Av}

\begin{deluxetable}{lcc}
\tablecaption{The Best-fit Parameters and Properties for o006\_s35616 with Extinction Parameter\label{tab:Av}}
\setlength{\tabcolsep}{0.25cm}
\tablehead{\colhead{} & \multicolumn{2}{c}{o006\_s35616}\\
\colhead{} & \colhead{Sonora Elf Owl} & \colhead{BT-settl CIFIST}} 
\startdata
$\log K_\mathrm{zz}$ [$\mathrm{cm^2\ s^{-1}}$] & $2.99^{+1.32}_{-0.71}$ & * \\
$T_\mathrm{eff}$ (K) & $2285.98^{+8.81}_{-7.60}$ & $2455.77^{+3.71}_{-3.94}$ \\
$\log g$ [$\mathrm{cm\ s^{-2}}$] & $3.33^{+0.06}_{-0.05}$ & $4.24^{+0.04}_{-0.03}$ \\
$\mathrm{[M/H]}$ & $-0.71\pm 0.02$ & * \\
C/O & $0.66\pm 0.02$ & * \\
$A_\mathrm{V}$ (mag) & $3.08^{+0.05}_{-0.04}$ & $3.91\pm 0.02$ \\
$\log(R^2/D^2)$ & $-23.48\pm 0.01$ & $-23.56\pm 0.01$ \\
$M/M_\mathrm{Jup}$ & $<8.4$ & $21.7^{+1.0}_{-0.7}$ \\
Age (Myr) & $<1$ & $18.6^{+1.4}_{-0.7}$ \\
$R/R_\mathrm{Jup}$ & $>2.7$ & $1.8\pm 0.1$ \\
Distance (kpc) & $>3.4$ & $2.5\pm 0.2$ \\
\enddata

\tablecomments{Similar to Table \ref{tab:parameter}, but incorporating the extinction parameter in the fitting process.}
\end{deluxetable}

\begin{figure*}[t]
\centering 
\includegraphics[width=1\textwidth]{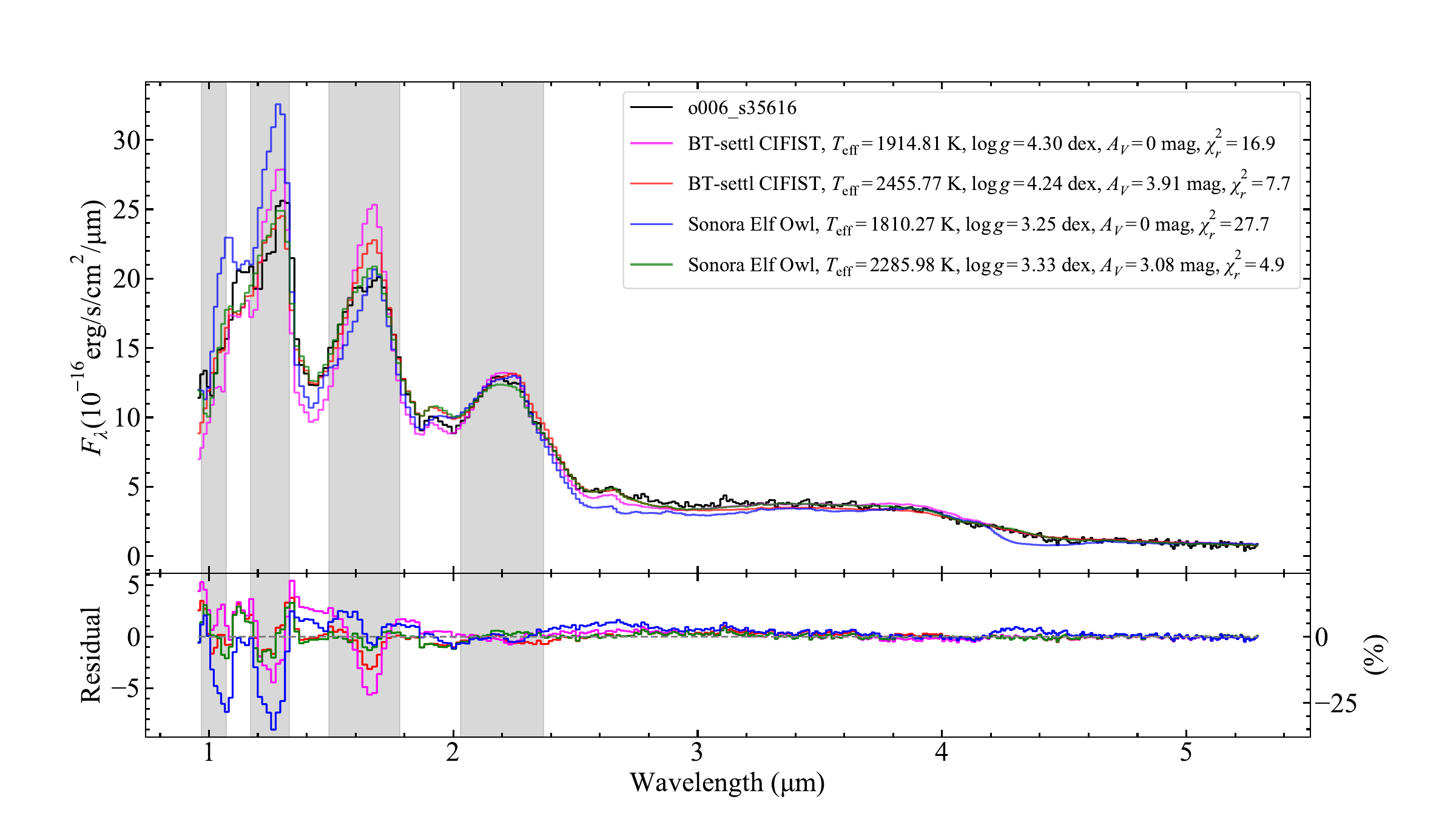}
\caption{Comparisons of the observed spectrum of o006\_s35616 with the best-fit model spectra obtained both with and without including interstellar extinction parameter. The gray-shaded regions indicate the wavelength intervals corresponding to the $Y$, $J$, $H$, and $K$ photometric bands.\\\label{fig:Avcomparison}} 
\end{figure*}

Since there are still significant residuals ($\sim20\%$ of the maximum observed flux) between the best-fit model spectra and observed spectra in the $Y$, $J$, and $H$ bands for the two L dwarf candidates, particularly for o006\_s35616, we explore whether incorporating the reddening effect caused by dust can enhance the accuracy of the model fits and the reliability of the derived parameters.

We used the method described in Section \ref{fit method}, introducing an additional parameter for extinction ($A_\mathrm{V}$). The extinction law ($A_\lambda/A_\mathrm{V}$) was adopted from \citet{2024ApJ...964L...3W}, which provides the interstellar dust extinction law for the JWST NIRSpec wavelength range of 0.6--5.3 $\mu$m. This extinction law is also consistent with the widely used Galactic average extinction law of \citet{2019ApJ...877..116W}. Given that the extinction law of \citet{2024ApJ...964L...3W} was determined from a sample of M/L dwarf candidates and closely matches the wavelength range of our spectra, it is particularly appropriate for our sources. 

It is important to note that the extinction law was used to simulate the effect of dust grains in an object’s atmosphere on its synthetic spectra, rather than to account for actual interstellar extinction. As discussed in Section \ref{fit method}, checks of the 3D extinction map for the three sources indicate that their interstellar extinctions are negligible. 

The fitting results incorporating extinction parameter reveal that for the cooler T-type brown dwarf candidate, o006\_s00089, the fitted value of $ A_\mathrm{V} $ converges to zero, and all other physical parameters remain consistent with the results from fits without extinction. Thus, introducing extinction has no significant impact on the fitting results for this source. 

In contrast, for o005\_s41280, the fitted $A_\mathrm{V}$ lies in the range of 1--1.5 mag. Including $ A_\mathrm{V} $ as a free parameter results in an increase in the $T_\mathrm{eff}$ by approximately 200 K for both models and a slight increase in $ \log g $. However, the best-fit model spectra with and without $A_\mathrm{V}$ exhibit no substantial differences.

Special attention should be given to the parameter changes observed for o006\_s35616 after introducing $ A_\mathrm{V} $. The best-fit parameters and properties are summarized in Table \ref{tab:Av}, and corresponding model spectra are presented in Figure \ref{fig:Avcomparison}. In this figure, the pink and blue spectra correspond to the results without $ A_\mathrm{V} $, while the red and green spectra represent the BT-settl CIFIST and Sonora Elf Owl models, respectively, with $ A_\mathrm{V} $ included. For o006\_s35616, the fitted $ A_\mathrm{V} $ values range from 3 to 4 mag for both models. Incorporating $ A_\mathrm{V} $ significantly reduces the residuals between the observed and model spectra across the $Y$, $J$, $H$, and other bands, leading to a substantial decrease in the reduced $\chi^2$ values.

However, we also note the dramatic changes in $ T_{\mathrm{eff}} $. For the Sonora Elf Owl model, the inclusion of $ A_\mathrm{V} $ increases $ T_{\mathrm{eff}} $ by approximately 475 K, while for the BT-settl CIFIST model, it increases by over 500 K. This large shift is consistent with the findings of \citet[][see Figure 25]{2024ApJ...961..121H}. They also pointed out that introducing the extinction for hotter L-type brown dwarfs may yield unrealistic physical parameters, such as $ T_{\mathrm{eff}} $. This discrepancy likely arises because the interstellar extinction model cannot adequately account for the dust properties in L-type brown dwarfs, as the dust in their atmospheres differs significantly from interstellar dust in terms of grain size, spatial distribution, and chemical composition. This issue has been corroborated by the work of \citet{2014MNRAS.439..372M} and \citet{2016ApJ...830...96H}, who demonstrated that due to limitations in modeling dust properties and distribution, current substellar atmospheric models often struggle to accurately simulate L dwarfs in these bands. 

Our results, consistent with those of \citet{2024ApJ...961..121H}, further highlight the limitations of current substellar atmospheric models in accounting for dust opacity for hotter L-type dwarfs. This could explain the observed discrepancies between model fits and observed spectra, suggesting that the treatment of dust in these models requires further refinement.

\section{Discussion} \label{sec:discussion}

In this section, we will discuss the observed colors of these three brown dwarf candidates with a focus on how such objects can be detected in current and future deep extragalactic catalogs. Additionally, we will compare the physical parameters of the brown dwarf candidate o006\_s00089, which was also identified by \citet{2024ApJ...964...66H} and designated as CEERS-EGS-BD-4.

\subsection{Comparing to the Colors of LRDs and High-Redshift Galaxies}

\begin{figure*}[t]
\centering 
\includegraphics[width=0.5\textwidth,height=0.4\textheight]{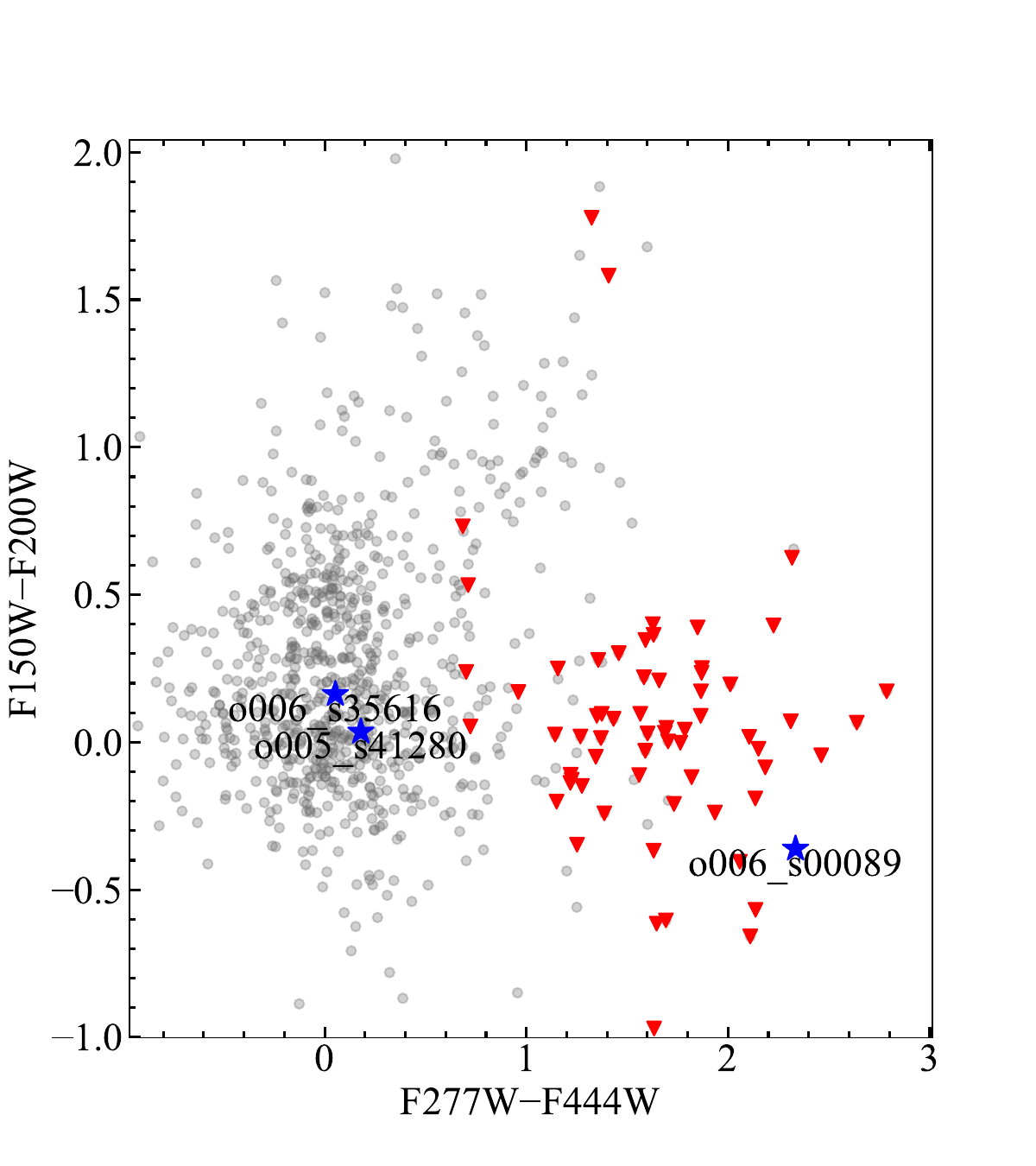}\includegraphics[width=0.5\textwidth,height=0.4\textheight]{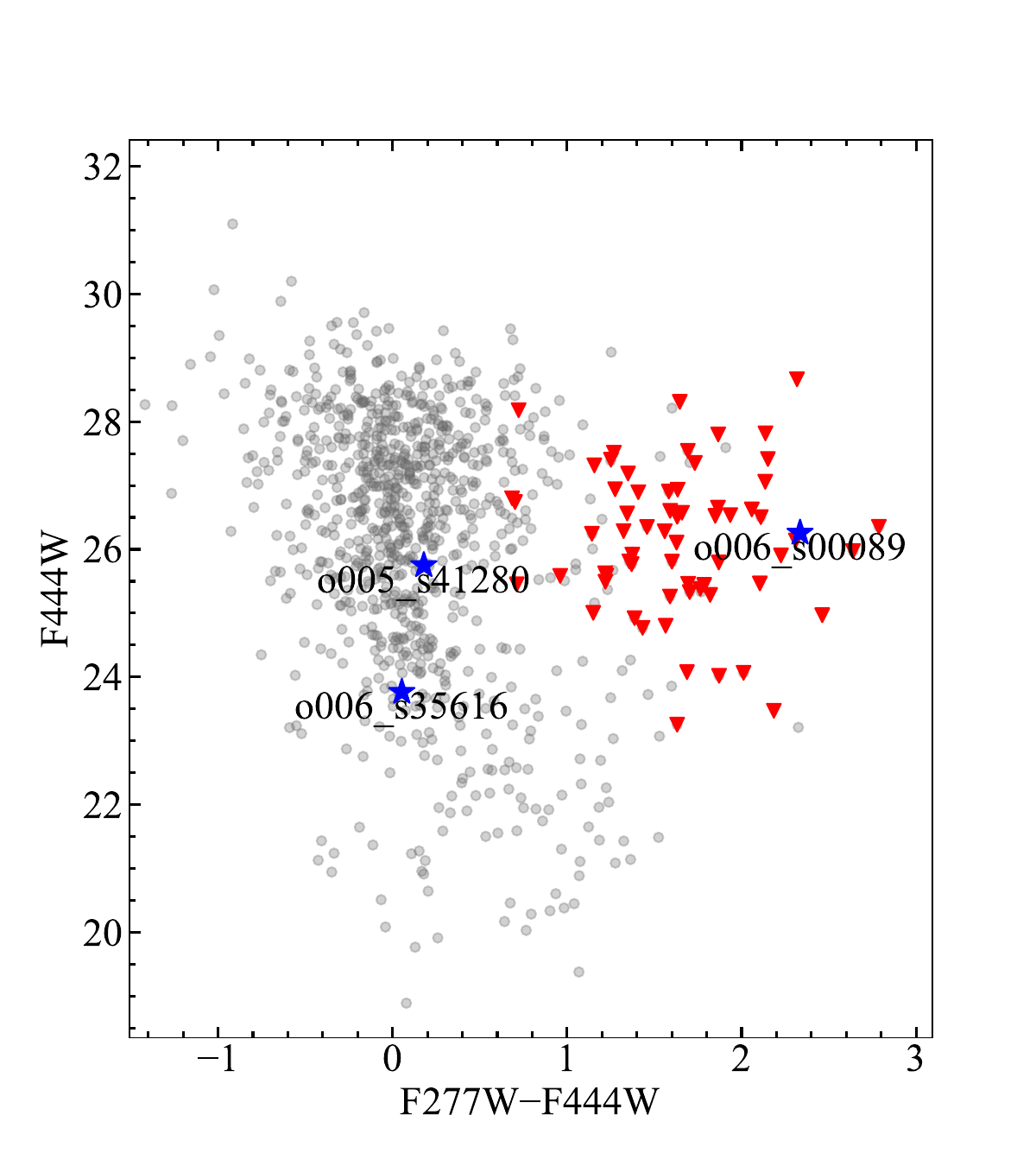}
\caption{Distribution of the three brown dwarf candidates (blue stars), sources from RUBIES (gray points), and LRDs (red triangles) on the color–color and color-magnitude diagrams. The LRDs are drawn from the CEERS catalog by \citet{2024arXiv240403576K}, and all photometric values are provided by \citet{2024AA...691A.240M}.\\\label{fig:CCD}} 
\end{figure*}

One of the key challenges in deep extragalactic surveys is distinguishing brown dwarfs from LRDs and high-redshift galaxies due to their overlapping positions in certain color-magnitude spaces. LRDs typically exhibit $ \mathrm{F277W - F444W} $ colors greater than 1 mag, whereas high-redshift galaxy candidates tend to occupy a broader region in the color-magnitude diagram due to variations in redshift, dust extinction, and star formation rates. This contrast provides a basis for disentangling the high-redshift galaxies and LRDs. However, the potential for misclassification remains, particularly between LRDs, high-redshift galaxies, and brown dwarfs. For instance, the cooler T and Y dwarfs like o006\_s00089, can mimic the $ \mathrm{F277W - F444W} $ colors of LRDs due to their similarly red colors caused by strong molecular absorption bands in the near-infrared band \citep[e.g.,][]{2023ApJ...957L..27L,2024ApJ...968....4P}.

To investigate the distributions of the three brown dwarf candidates, all sources from RUBIES (mostly high-redshift galaxy candidates), and LRDs on the color-color diagram (CCD) and color-magnitude diagram(CMD), we cross-matched the coordinates of all sources from RUBIES with those of the CEERS sources provided by \citet{2024AA...691A.240M} to obtain the photometric data. We also did a similar cross-match using the LRD coordinates from \citet{2024arXiv240403576K}, resulting in diagrams of $ \mathrm{F277W - F444W} $ colors versus $ \mathrm{F150M - F200W} $ colors, and $ \mathrm{F277W - F444W} $ colors versus the $ \mathrm{F444W} $ magnitude.  

As shown in Figure~\ref{fig:CCD}, we found that LRDs, with their redder $ \mathrm{F277W - F444W} $ colors, are distinctly separated from the cluster of high-redshift galaxies. However, o006\_s00089 overlaps with these LRDs, making it difficult to distinguish o006\_s00089 from the LRDs on both the CCD and CMD. Its overlap with LRDs suggests that cooler brown dwarfs are more likely to be mistaken for LRDs, reinforcing the importance of accurate model-based fitting to resolve ambiguities.

On the other hand, the hotter L dwarf candidates, o005\_s41280 and o006\_s35616, lie within the cluster of high-redshift galaxies in RUBIES, contaminating with these objects. The relatively bluer colors of the two sources suggest that hotter brown dwarfs are less likely to overlap with the regions occupied by LRDs or cooler T and Y dwarfs, but highlight the complexity of distinguishing between hotter ultracool dwarfs (M-L types) and high-redshift galaxies. Therefore, selecting redder sources, as in the method used by \citet{2024ApJ...964...66H} to identify brown dwarf candidates (\(m_\mathrm{F444W, AB} \leq 28.5\) mag and \(\mathrm{F277W - F444W} \geq 0\) mag), tends to exclude hotter L dwarfs, thereby only identifying the later-type T or Y dwarfs. 

We also tried to utilize F115W, F356W, and the above photometric data, combining them to form numerous new CCDs, but none of these combinations could simultaneously distinguish the brown dwarf candidates, the high-redshift galaxy cluster, and the LRDs.

Based on the observed distribution in the CCD and CMD, and the total number of NIRSpec PRISM/CLEAR spectra we derived in RUBIES, we estimate a brown dwarf contamination rate of approximately 0.1\% (3/3194) in extragalactic deep field surveys. The likelihood of identifying relatively hotter M-L type dwarfs (such as o005\_s41280 and o006\_s35616) is higher than that of detecting cooler T or Y dwarfs (like o006\_s00089). However, it is important to note that the sample we used in RUBIES represents a highly specific subset of the CEERS and PRIMER fields, selected based on photometric properties and brightness (e.g., bright and red sources). This selection inherently favors brown dwarfs and likely inflates the contamination rate.

In comparison, \citet{2024ApJ...964...66H} identified 7 brown dwarf candidates among over 80,000 photometric sources in the CEERS field (\(7/80000 \approx 0.009\%\)). However, their study focused only on late-type T and Y dwarfs and did not include L dwarfs, which may lead to an underestimation of the contamination rate. Considering these differences in sample selection and analysis focus, we estimate that the true brown dwarf contamination rate in extragalactic deep fields lies between \(\sim0.01\%\) and \(0.1\%\). A more precise contamination rate can be determined by combining and re-evaluating results from both approaches once additional extragalactic deep field spectroscopic data become publicly available.

To further refine the detection and classification of brown dwarfs in extragalactic fields, future surveys could benefit from expanding the wavelength coverage into the mid-IR, where brown dwarfs exhibit distinct spectral features compared to galaxies.

\subsection{Comparison of Parameters of o006\_s00089 with Literature}

The brown dwarf candidate o006\_s00089 was also identified in the sample of \citet{2024ApJ...964...66H}, where it was designated as CEERS-EGS-BD-4. Their selection process relied on photometric data, applying color and magnitude thresholds, such as $m_\mathrm{F444W, AB} \leq 28.5$ mag and $\mathrm{F277W - F444W}  \geq 0$ mag, to isolate brown dwarf candidates. These candidates were further confirmed using a $\chi^2$ minimization method. Physical parameters were then derived by fitting three atmospheric models--—Sonora Cholla, Sonora Bobcat, and ATMO2020. The evolutionary properties, including distances, were calculated using the Sonora Bobcat evolutionary models.

A detailed comparison between our results and those obtained by \citet{2024ApJ...964...66H} reveals a general consistency in the derived physical parameters of the brown dwarf candidate o006\_s00089. The effective temperatures are closely aligned, with $T_\mathrm{{eff, Cholla}} = 1050 \, \mathrm{K}$, $T_\mathrm{{eff, Bobcat}} = 1000 \, \mathrm{K}$, and $T_\mathrm{{eff, ATMO2020}} = 1000 \, \mathrm{K}$. These values agree well with our derived temperatures, further supporting the classification of this object as a late-type T dwarf.  

The surface gravities obtained in their study ($\log g = 5.0 \, [\mathrm{cm\,s^{-2}}]$ from all three models) fall within the range of our results, $\log g = 4.34 - 5.5 \, [\mathrm{cm\,s^{-2}}]$. This suggests general consistency across different methods and atmospheric models, despite potential differences in model assumptions and fitting techniques.  

Their study reported radii of $0.92 \, R_\mathrm{Jup}$ for both the Sonora Cholla and Sonora Bobcat models, and $1.27 \, R_\mathrm{Jup}$ for the ATMO2020 model, with corresponding distances of 1391, 1350, and 1813 pc, respectively. Using our fitted parameters ($T_\mathrm{eff}$, $\log g$, and $\log(R^2/D^2)$), we derived $R$ of less than $1.2 \, R_\mathrm{Jup}$ and distances within 2 kpc, which are consistent with their results. Furthermore, by assuming an age range of 1--5 Gyr, we estimated radii of $0.8 - 0.9 \, R_\mathrm{Jup}$ and distances of $1.4 - 1.7 \, \mathrm{kpc}$, which also generally agree with their findings.

Furthermore, this consistency across independent analyses highlights the reliability of using photometric data or spectroscopic methods to characterize brown dwarfs in extragalactic deep field surveys. It also emphasizes the importance of multi-model comparisons in refining parameter estimates. The slight variations in surface gravity, radius, and distance values may reflect differences in model assumptions, such as metallicity, cloud treatment, or dust opacity, pointing to the need for ongoing improvements in brown dwarf atmospheric and evolutionary models.

\section{Conclusion} \label{sec:conclusion}

In this study, with JWST NIRSpec PRISM/CLEAR spectra from RUBIES, we found three brown dwarf candidates, o006\_s00089, o005\_s41280, and o006\_s35616. We further applied a Bayesian framework with nested sampling algorithm to fit the physical parameters of the three brown dwarf candidates using updated atmospheric models, including Sonora Elf Owl, ATMO2020++, and BT-settl CIFIST. The evolutionary properties, such as mass, age, radius, and distances were derived using the evolutionary model from \citet{Chabrier2023AA}. The main results of this work are as follows.

\begin{enumerate}
    \item The fitting results indicate that the $T_\mathrm{eff}$ of o005\_s41280 and o006\_s35616 are within the ranges of 2100-2300 K and 1800-2000 K, suggesting they are likely L dwarfs. In contrast, o006\_s00089 has an $T_\mathrm{eff}$ below 1000 K, classifying it as a late T dwarf.
    \item Current atmospheric models may struggle to accurately constrain the $\log g$ of L dwarfs, resulting in significant discrepancies in $\log g$ values derived from different models. These variations propagate to other physical properties, such as mass, age, radius, and distance, making them challenging to constrain with high precision. Our analyses through various models and methods suggest that the distances of the three brown dwarf candidates are basically around 2 kpc. Furthermore, substantial differences are observed between the best-fit model spectra and the observed spectra of L dwarfs, particularly in the $Y$, $J$, and $H$ bands. These discrepancies are likely attributed to insufficient dust content included in the atmospheric models for M/L dwarfs.
    \item The positions of the three candidate brown dwarfs on the diagrams of $ \mathrm{F277W - F444W} $ colors versus $ \mathrm{F150M - F200W} $ colors, and $ \mathrm{F277W - F444W} $ colors versus the $ \mathrm{F444W} $ magnitude, illustrate significant relationships between brown dwarfs, high-redshift galaxies, and LRDs. The cooler source, o006\_s00089, overlaps with the region mainly occupied by LRDs, whereas the hotter sources, o005\_s41280 and o006\_s35616, are situated within the high-redshift galaxy cluster. This highlights the challenges in distinguishing these populations based solely on these near-infrared photometric data. According to our analysis, we estimate a brown dwarf contamination rate of approximately 0.1\% in extragalactic deep field surveys, with a higher detection probability for M-L type ultracool dwarfs compared to cooler T and Y dwarfs.
    \item Comparisons with the results of \citet{2024ApJ...964...66H} for o006\_s00089 show overall consistency in derived parameters, including $T_\mathrm{eff}$, $\log g$, radius, and distance. This agreement further validates the physical parameters and properties of this source. Additionally, these findings emphasize the reliability of both photometric and spectroscopic methods in effectively characterizing brown dwarfs in extragalactic deep field surveys.
\end{enumerate}

Our study demonstrates the feasibility of identifying brown dwarf candidates in extragalactic deep field surveys through spectroscopic methods combined with atmospheric models, and contributes to a deeper understanding of the characteristics of brown dwarfs and their potential contamination in deep-field catalogs. Additionally, these findings underscore the importance of developing alternative color or magnitude selection techniques to effectively distinguish brown dwarfs from high-redshift galaxies and LRDs, improving the reliability of such classifications in future surveys.

\section*{Acknowledgments} 
We thank the anonymous referee for helpful comments/suggestions. This work is supported by the National Natural Science Foundation of China (NSFC) through the projects 12373028, 12322306, 12173047, 12133002, 11988101, and 11933004. S. W. and X. C. acknowledge support from the Youth Innovation Promotion Association of the CAS with Nos. 2023065 and 2023055. This work is based on observations made with the NASA/ESA/CSA James Webb Space Telescope. The data were obtained from the Mikulski Archive for Space Telescopes (MAST) at the Space Telescope Science Institute.

\software{NumPy \citep{numpy}, Matplotlib \citep{matplotlib}, corner.py \citep{corner}, UltraNest \citep{ultranest}}

\appendix
\counterwithin{figure}{section}
\section{Corner Plots of the Brown Dwarf Candidates}

\begin{figure*}
\centering 
\includegraphics[width=1.\textwidth]{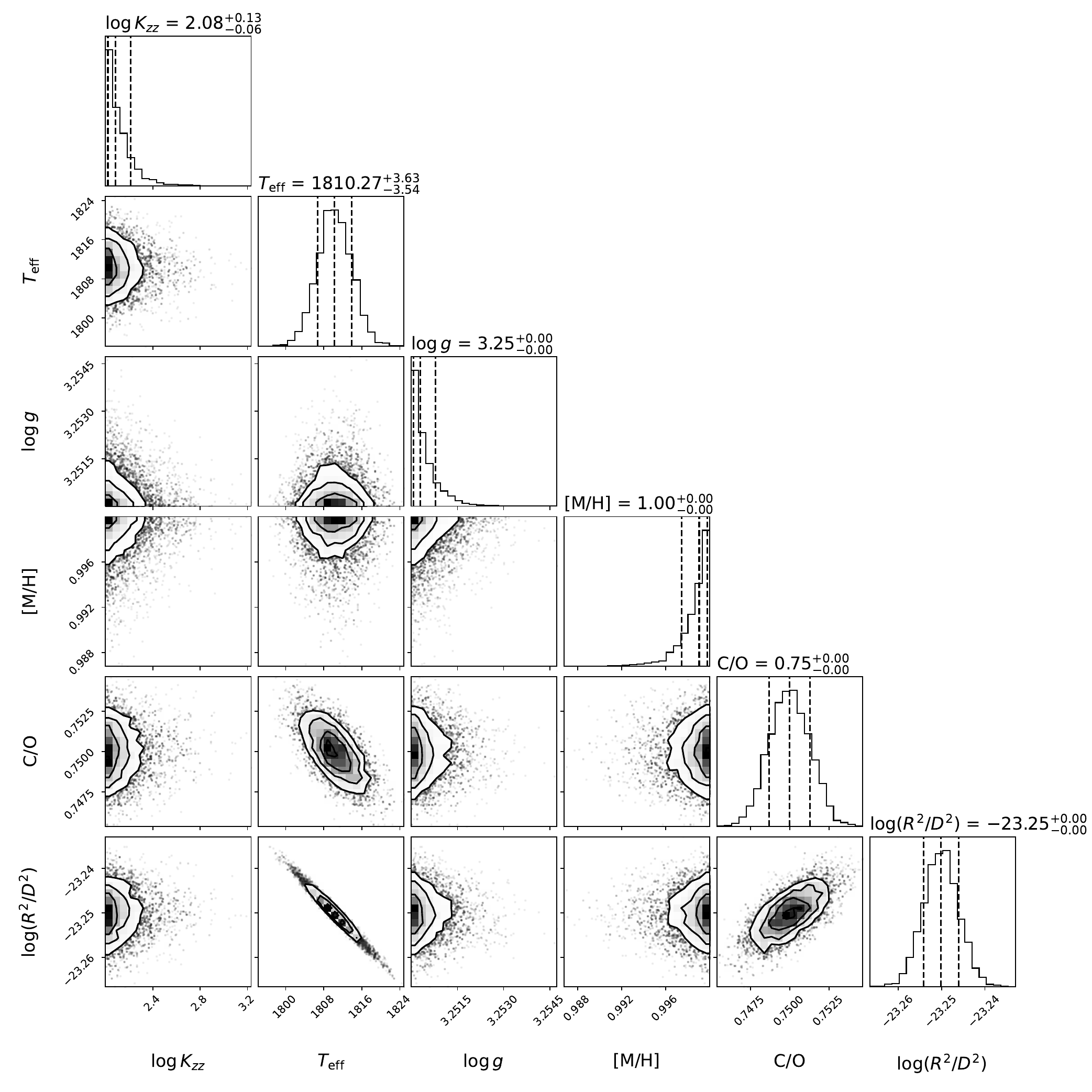}
\caption{The marginalized posterior distribution of six parameters for the observed spectra of o006\_s35616, using the Sonora Elf Owl models. The parameters are, in order, the logarithm of vertical eddy diffusion parameter $\log K_\mathrm{zz}$, effective temperature $T_\mathrm{eff}$, surface gravity $\log g$, metallicity [M/H], carbon–to–oxygen ratio C/O, and scaling factor $\log(R^2/D^2)$. \label{fig:corner_owl_35616}} 
\end{figure*}

\begin{figure*}
\centering 
\includegraphics[width=1.\textwidth]{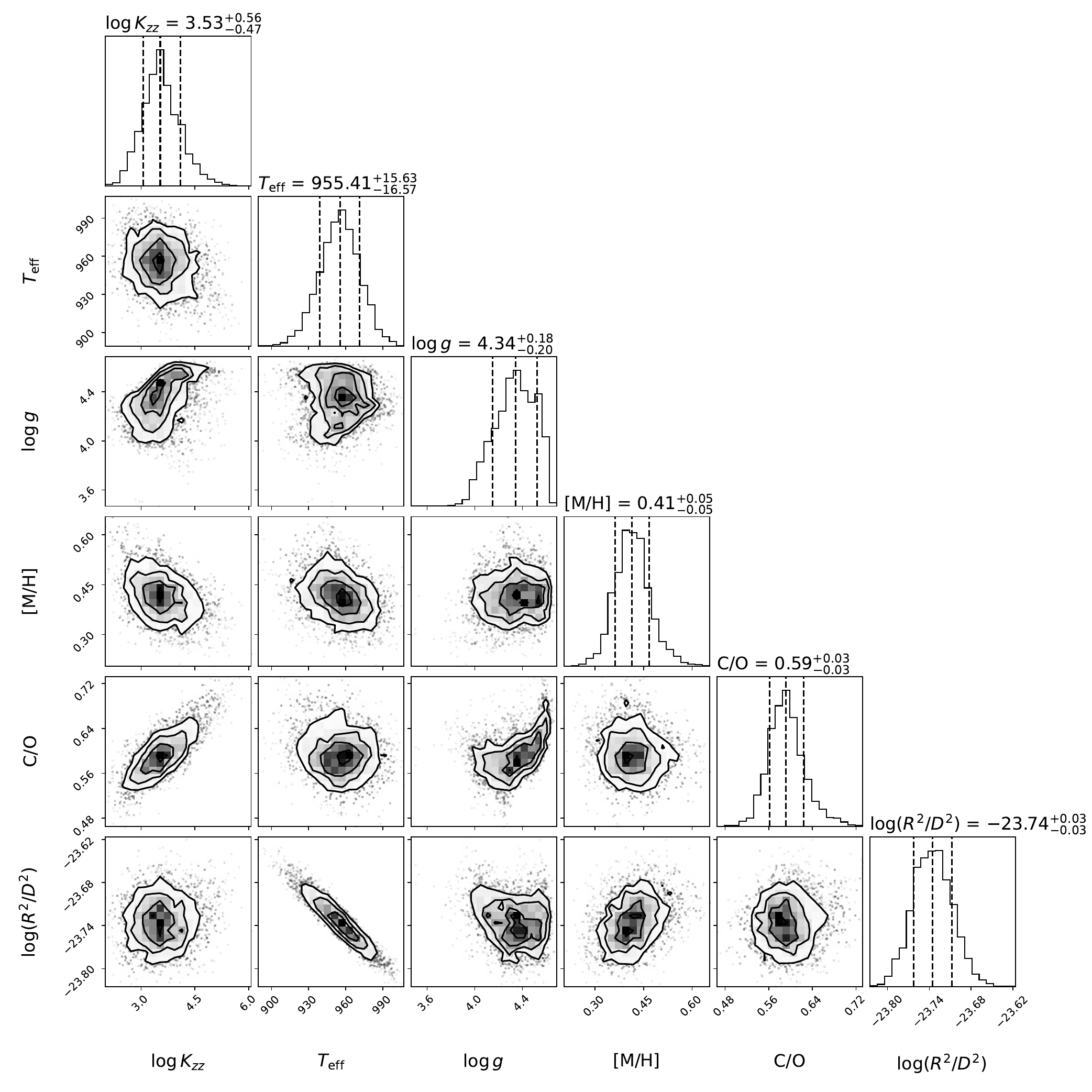}
\caption{Similar to Figure~\ref{fig:corner_owl_35616}, but for o006\_s00089 using the Sonora Elf Owl models.\label{fig:corner_owl_89}} 
\end{figure*}

\begin{figure*}
\centering 
\includegraphics[width=1.\textwidth]{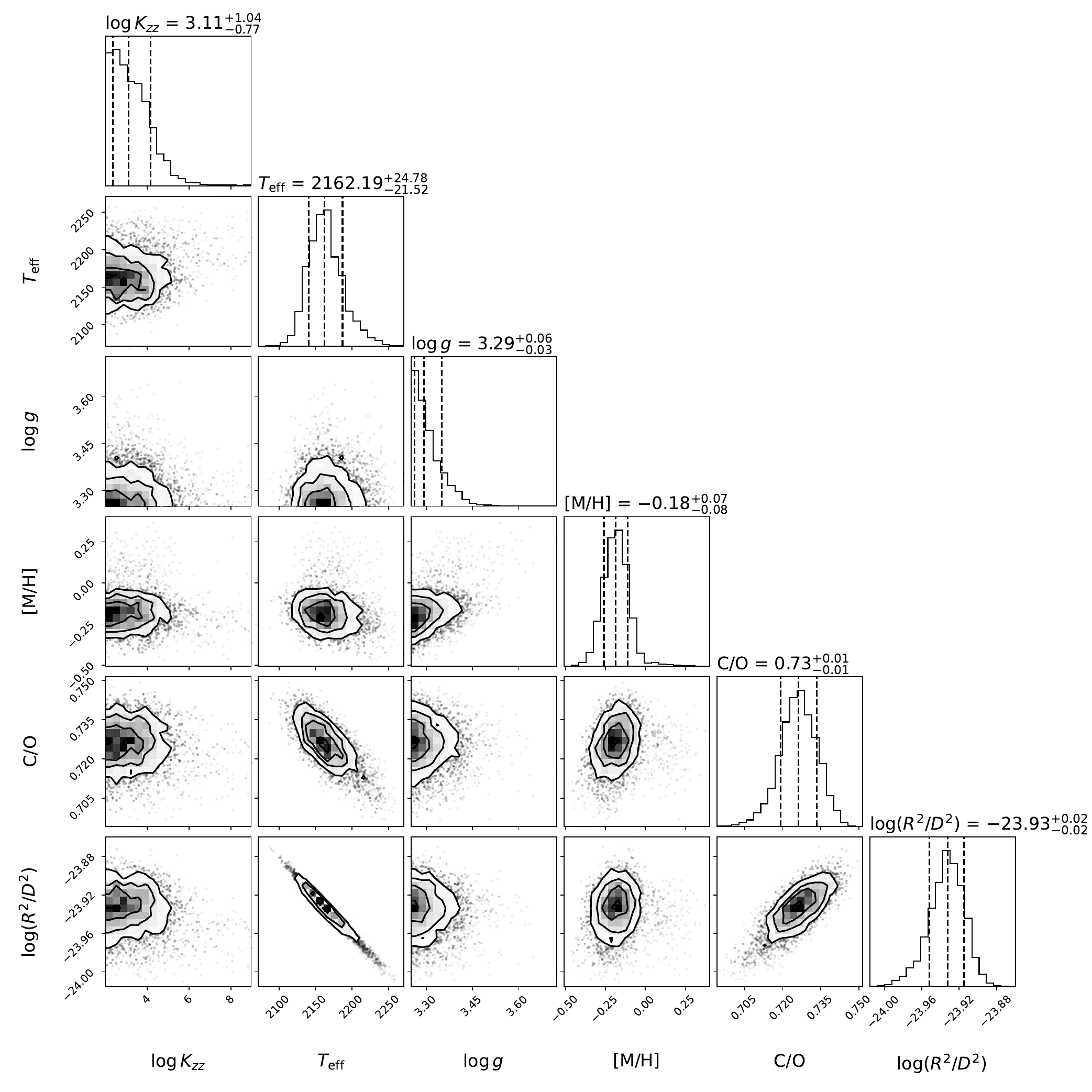}
\caption{Similar to Figure~\ref{fig:corner_owl_35616}, but for o005\_s41280 using the Sonora Elf Owl models.\label{fig:corner_owl_41280}} 
\end{figure*}

\begin{figure*}
\centering 
\includegraphics[width=1.\textwidth]{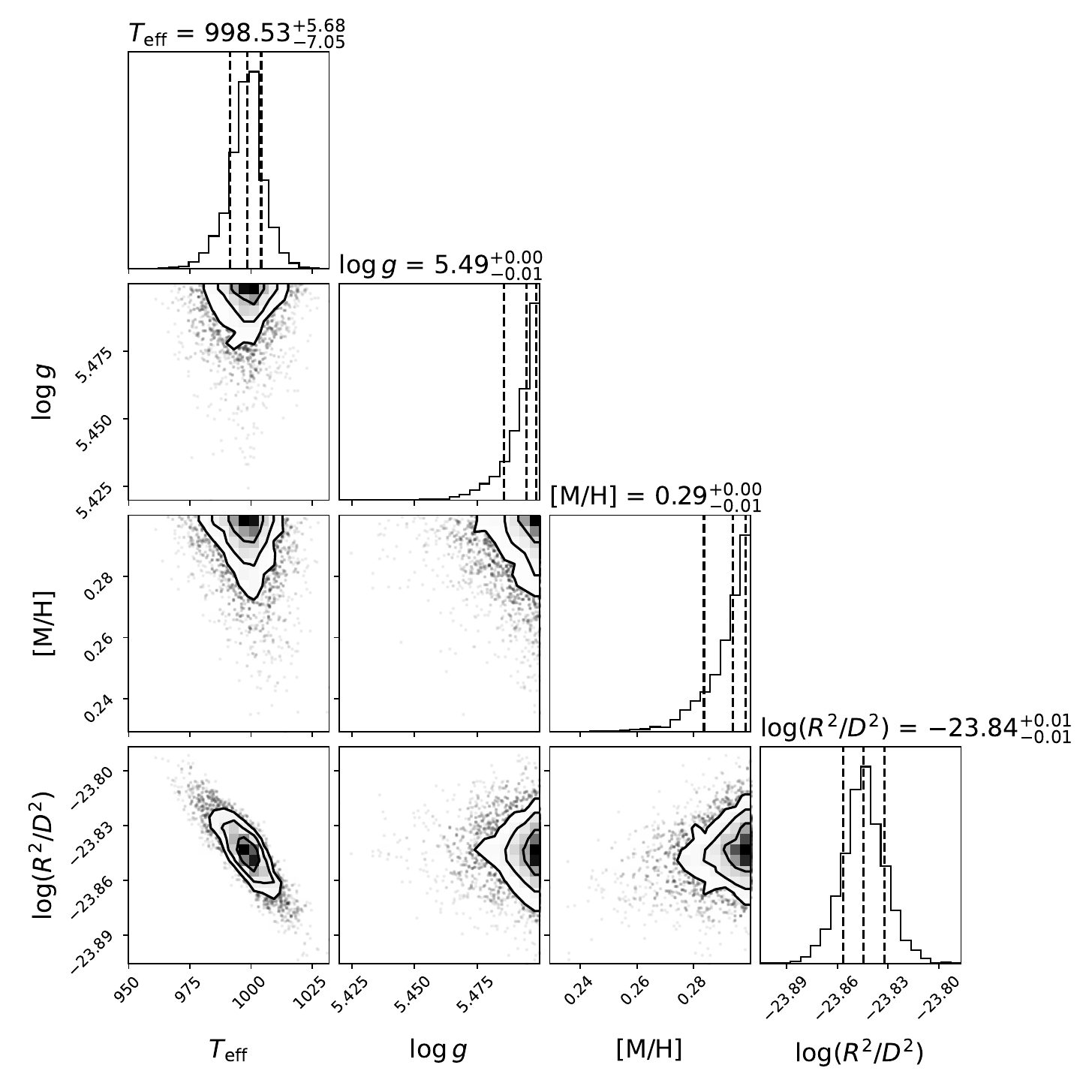}
\caption{The marginalized posterior distribution of four parameters for the observed spectra of o006\_s00089, using the ATMO2020++ models. The parameters are, in order, effective temperature $T_\mathrm{eff}$, surface gravity $\log g$, metallicity [M/H], and scaling factor $\log(R^2/D^2)$. \label{fig:corner_ATMO_89}} 
\end{figure*}

\begin{figure*}
\centering 
\includegraphics[width=.5\textwidth]{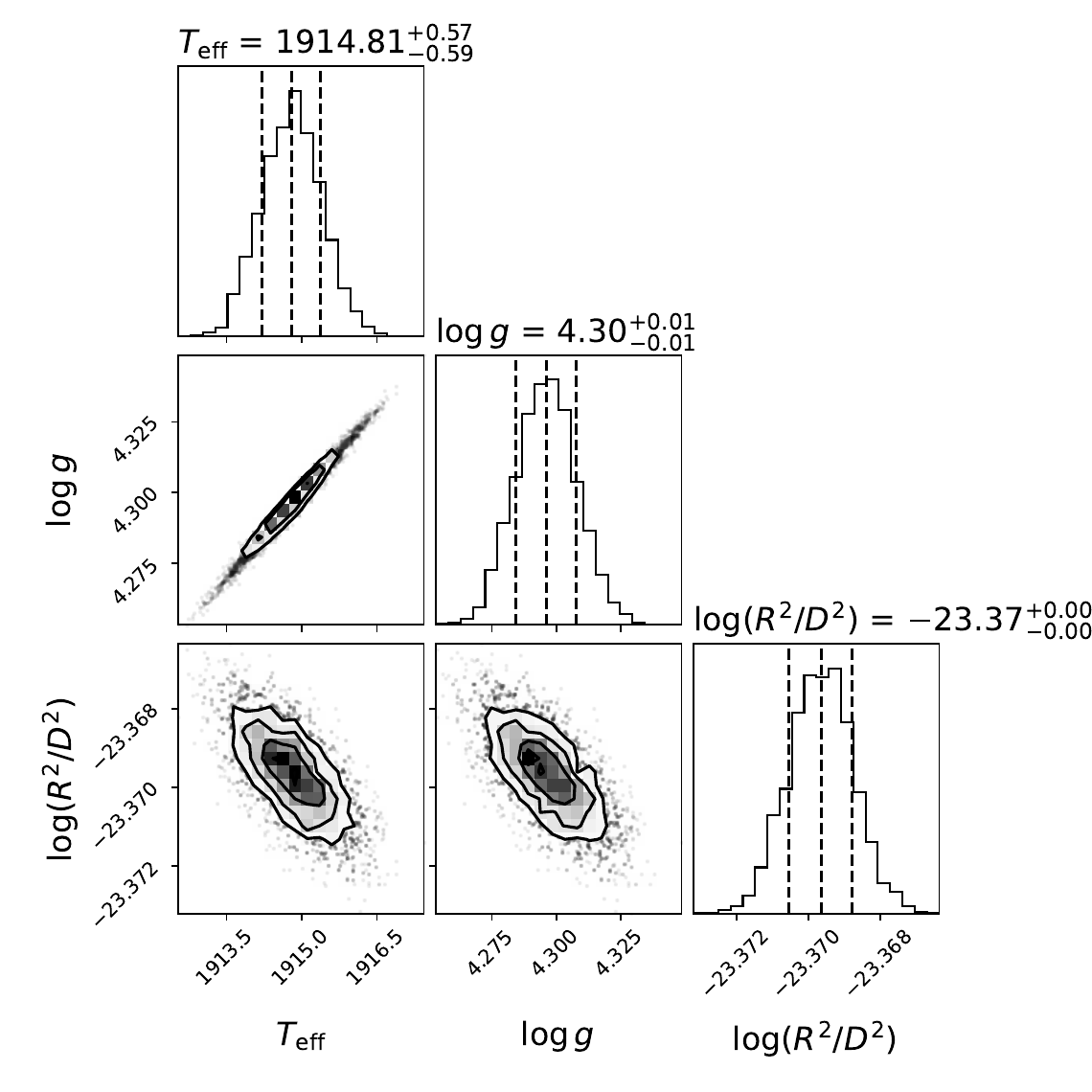}\includegraphics[width=.5\textwidth]{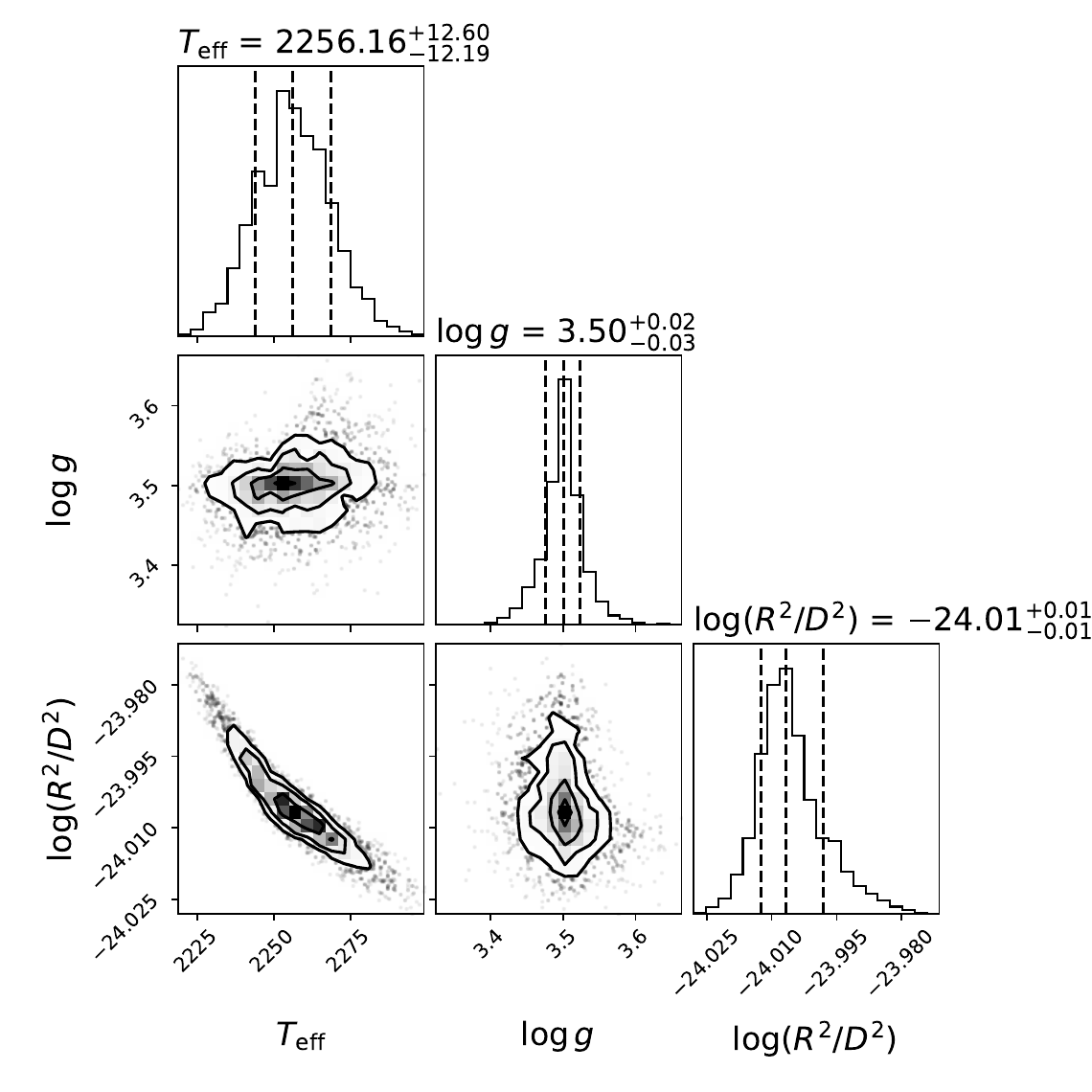}
\caption{The marginalized posterior distribution of three parameters for the observed spectra of o006\_s35616 (left) and o005\_s41280 (right), using the BT-settl CIFIST models. The parameters are, in order, effective temperature $T_\mathrm{eff}$, surface gravity $\log g$, and scaling factor $\log(R^2/D^2)$. \label{fig:corner_BT_35616}} 
\end{figure*}

\bibliography{main}{}
\bibliographystyle{aasjournal}

\end{document}